\documentclass[12pt]{article}
\usepackage{fullpage,setspace}
\usepackage{amssymb, amsthm, amsmath}
\usepackage{graphicx}
\usepackage[authoryear]{natbib}
\usepackage{bm}
\usepackage{lineno}
\usepackage{times}
\usepackage{url}

\usepackage{xcolor}
\usepackage{graphicx}
\usepackage{amsmath}
\usepackage{makecell}
\usepackage{amssymb}
\usepackage{url}

\newcommand{\bR}{ \mbox{\bf R}}

\newcommand{\bU}{ \mbox{\bf U}}
\newcommand{\bW}{ \mbox{\bf W}}
\newcommand{\bbb}{\boldsymbol}

\newcommand{\indep}{\rotatebox[origin=c]{90}{$\models$}}

\usepackage{xcolor}
\usepackage{float}
\usepackage{csvsimple}
\usepackage{adjustbox}
\usepackage{makecell}
\usepackage{tikz}
 
\allowdisplaybreaks
\usepackage{tipa}

\newtheorem{assumption}{Assumption}
\newtheorem{theorem}{Theorem}

\begin{document} % \linenumbers

\begin{center}
{\Large Estimating the Causal Effect of Redlining on Present-day Air Pollution}\\\vspace{6pt}
{\large Xiaodan Zhou\footnote[1]{North Carolina State University}, Shu Yang$^1$, Brian J Reich$^1$}\\
% \today
\end{center}

\begin{abstract}\begin{singlespace}\noindent

Recent studies have shown associations between redlining policies (1935-1974) and present-day fine particulate matter (PM$_{2.5}$) and nitrogen dioxide (NO$_2$) air pollution concentrations. In this paper, we reevaluate these associations using spatial causal inference.  
Redlining policies enacted in the 1930s, so there is very limited documentation of pre-treatment covariates. Consequently, traditional methods fails to sufficiently account for unmeasured confounders, potentially biasing causal interpretations. By integrating historical redlining data with 2010 PM$_{2.5}$ and NO$_2$ concentrations, our study aims to discern whether a causal link exists. Our study addresses challenges with a novel spatial and non-spatial latent factor framework, using the unemployment rate, house rent and percentage of Black population in 1940 U.S. Census as proxies to reconstruct pre-treatment latent socio-economic status. We establish identification of a causal effect under broad assumptions, and use Bayesian Markov Chain Monte Carlo to quantify uncertainty. 
Our analysis indicates that historically redlined neighborhoods are exposed to notably higher NO$_2$ concentration. In contrast, the disparities in PM$_{2.5}$ between these neighborhoods are less pronounced. Among the cities analyzed, Los Angeles, CA, and Atlanta, GA, demonstrate the most significant effects for both NO$_2$ and PM$_{2.5}$. 

\vspace{12pt}
\end{singlespace}\end{abstract}

\section{Introduction}\label{s:intro}

\subsection{Redlining policy}

The redlining policy, initiated in 1935 by the Federal Home Loan Bank Board, mandated the Home Owners' Loan Corporation (HOLC) to create `residential security maps’. These maps classified residential regions with grades reflecting investment security: `A' for Desirable, `B' for Still Desirable, `C' for Declining, and `D' for Redlined. This grading system, operational until the 1974, directly influenced lending decisions. Regions graded `A' were considered minimal risk by banks and mortgage lenders for loans and safe investments, while those labeled `D' were deemed hazardous.

Some studies have investigated the financial inequalities stemming from the redlining policy, with a emphasis on causal analysis and addressing potential confounders. \cite{aaronson2021effects} employed a boundary-design approach to mitigate these confounders. Their analysis concentrated on areas adjacent to redlining boundaries (D), comparing `treated' and `controlled' boundaries using propensity score weighting. They discovered that regions assigned to be redlined experienced deteriorating housing market outcomes in the following decades. Similarly, \cite{fishback2020holc} conducted a detailed boundary analysis, examining socio-economic characteristics near C-D grade borders. They observed a decline in home values and an increase in black population shares on the D-side compared to the C-side.

The growing interest in environmental inequality has led to association-based studies concerning the redlining policy. \cite{lane2022historical} revealed a consistent and nearly monotonic relationship between air pollutants and redlining grades, noting particularly an increase (over 50\%) in NO$_2$ concentrations from A-graded to D-graded regions. The study also found that % intra-urban disparities for NO$_2$ and PM$_{2.5}$ are significantly larger by HOLC grade than by race and ethnicity; 
within each grade, disparities in air pollution exposure based on race and ethnicity continue to exist. This underscores the racially discriminatory impact of redlining on communities. Additionally, \cite{jung2022effects} discovered that between 1998 and 2012, in New York City, schools in historically redlined regions saw smaller reductions in combustion-related air pollutants compared to others. However, the direct causal link between redlining policies and air pollution exposure remains uncertain, despite the apparent association.

% similar but only CA \citep{estien2024historical} 
% \citep{nardone2020associations} find significantly worse birth outcomes and health disparities comparing C to B, and comparing C to D. 

In this paper, we revisit the data and apply methods from spatial causal inference to determine if the link between redlining and air pollution persists after accounting for spatial dependence and confounding variables. We 
% Our research aims to bridge the gap in understanding the causal effect of the redlining policy on contemporary air pollution exposure. 
% The recent digitization of survey data for 4,140 regions in 'residential security maps' by the Mapping Inequality Project (\url{https://dsl.richmond.edu/panorama/redlining}) provides a unique opportunity for this analysis. 
link this historical redlining data with current air pollution concentrations, and assesses the potential long-term environmental effects of redlining policies (1935-1974) on present-day air pollution concentrations. 

This analysis faces two key challenges. First, both redlining grades and air pollution concentrations exhibit spatial patterns, which must be carefully considered. Second, there is a risk of unmeasured confounding factors, particularly socio-economic status, that could influence both the historical redlining grades and current air pollution concentrations. In the following section, we outline our approach to addressing these challenges.

\subsection{Spatial Causal Inference and Unmeasured Confounding}
Addressing unmeasured confounding has become a major topic in causal inference. A unmeasured confounder could introduce bias into the estimated effect and lead to incorrect conclusions about the true causal relationship. There are a wide range of methods to adjust for unmeasured confounding, such as instrumental variables \citep{bound1995problems}, negative controls \citep{lipsitch2010negative}, latent and proxy variables \citep{kuroki2014measurement}. These methods are not specifically designed for spatial data, but have been adopted to account for spatial unmeasured confounding in application studies, such as \cite{davis2021propensity}, 
%accommodate propensity score matching for multilevel spatial data to account for spatial unmeasured confounding, 
\cite{shao2022metro}, \cite{giffin2021instrumental}, \cite{haschka2020provision}, 
\cite{tustin2017associations}, and \cite{jerzak2023integrating}. 

Moreover, causal methods applied to complex spatial data have been drawing attention. A spatial confounder is a unmeasured confounder that contains spatial structure. The `blessing' of spatial confounder, compared with unstructured confounder, is that the spatial information may be used to capture some of the variability in the confounder, thus mitigating the bias \citep{gilbert2021approaches}. \cite{dupont2022spatial+} developed method named `spatial+', for cases when the treatment is spatially dependent but not fully determined by spatial location. A partial linear regression was used to adjust for spatial confounding. \cite{guan2023spectral} assumed a global-scale confounding (global relative to the treatment variable) and adjusted for confounding in the spatial domain by adding a spatially smoothed version of the treatment to the mean of the response variable. In spatial+ and the spectral adjustment, the treatment was assumed continuous. Other methods include region adjustment via spatial smoothing \citep{schnell2020mitigating}, % provide a set of assumptions (given in the supporting in- formation) to identify the unmeasured confounding bias... if the spatial scale of treatment is larger or about the same as the unmeasured confounder, the confounding bias cannot be mitigated. 
distance adjusted propensity score matching \citep{papadogeorgou2019adjusting}, spatial propensity-score \citep{davis2019addressing}, which have been reviewed in \cite{reich2021review}. 

There are challenges in the redlining data that cannot be addressed by existing methods. Social-economic status is arguably the most important confounding variable. Though some relevant data can be found in the U.S. Census, it is dangerous to assume that we can use them to fully account for social-economic status, thus potentially biasing causal interpretations. Moreover, the time lapse of 75 years between the policy action and the pollution measurement further obscures causal links. In addition, the existence of spatial correlation in all of treatment, outcome, proxies, and potentially the latent confounder, complicates the problem. 

In response to these challenges, we expect a method that accommodates for latent confounding factors using proxy variables, and accommodates for unmeasured spatial confounders. We want such a method to sufficiently account for unmeasured confounders and draw consistent estimates. However, none of existing method would allow use the leverage the spatial structure and observed variable while taking use of the proxy variable for unobserved spatial confounding, while the use of proxy for non-spatial causal analysis has been prevalent such as \cite{Kong2019MulticauseCI}, % multi-cause
\cite{yang2020improved}, and %heterogeneous treatment
\cite{miao2018confounding}. %Confounding Bridge Approach
This paper aims to fill in the gap.

\subsection{Contributions and structure of the paper}
Our study develops a novel latent framework for causal inference that accounts for both spatial and non-spatial confounding. We establish the identification of causal effects under broad assumptions, and use Bayesian MCMC to quantify uncertainty. Our method promises to enhance the validity of causal claims by rigorously adjusting for confounders. In the case study, we assesses the potential long-term environmental effects of redlining policies on present-day air pollution concentrations. 

The remainder of the paper proceeds as follows. Section \ref{s:data} describes the motivating data. Sections \ref{s:method} and \ref{s:theory} detail the statistical methods and their theoretical properties. Section \ref{s:comp} discusses computational aspects. The method is assessed through a simulation study in Section \ref{s:sim}, and its application to the motivating data is presented in Section \ref{s:app}. The paper concludes with Section \ref{s:discussion}.

\section{Data description}\label{s:data}
The data for our study is drawn from multiple sources. We obtain the treatment variable, that is the redlining grades, from the Mapping Inequality Project \citep{nelson2023mapping}, including `A' for Desirable, `B' for Still Desirable, `C' for Declining, and `D' for Redlined. 
% We label the grades C and D as the `redlined', and grades A and B as the `non-redlined'. The details about this decision can be found in Appendix A.1. 

For the outcome variables, we use concentrations of annual-average values
fine particulate matter (PM$_{2.5}$) and nitrogen dioxide (NO$_2$) in 2010. This year is selected because comprehensive air pollution monitoring for PM$_{2.5}$ and NO$_2$ began in the late 1990s, with sufficient data becoming available from 2010 onward \citep{epaweb}. While 2020 pollution data is available, we prefer using 2010 as it is closer in time to when redlining policies were enacted, making it a more relevant measure for assessing long-term effects. The pollution data is derived from empirical models provided by the Center for Air, Climate, and Energy Solutions (CACES) \citep{kim2020concentrations}. To address potential confounding, we incorporate variables from the 1940 U.S. Census: unemployment rate, mean house rent, and percentage of Black population.

The geographical boundaries of the HOLC maps, the 1940 Census tracts, and the 2010 air pollution monitoring data differ. We merge these datasets by spatially overlapping them within the HOLC regions. Detailed methodologies for this spatial integration are available in Web Appendix A. After data cleaning, our final dataset includes 4,079 regions across 69 cities and 27 states in the U.S., covering about 20\% of the 1940 U.S. population. 

Web Table S1 presents key summary statistics from the 1940 Census and 2010 air pollution data comparing redlined and non-redlined groups across all cities. Example maps for Atlanta, GA are shown in Web Figure S1. There is clearly spatial dependence in the pollution, redlining and census variables. We observe higher mean values of NO$_2$, PM$_{2.5}$, unemployment rate, and percentage of Black population in areas with worse redlining grades. The percentage of Black population is zero-inflated, with approximately 5\% of the observed values being zero. Conversely, mean house rent follows an opposite trend, with higher rents observed in non-redlined areas. 

There is a substantial time gap between the redlining era (1935-1974) and the 2010 air pollution data. This introduces challenges such as potential attenuation of the redlining effects on air quality over time. Additionally, it complicates the identification and acquisition of confounders, particularly socio-economic status, which is a critical but debated concept among social scientists. To address this, we use data from the 1940 Census, including unemployment rates, housing conditions, and racial composition, as proxies for the underlying socio-economic status construct.

\section{Statistical methods}\label{s:method} 

The data are drawn from $M$ cities. City $i$ is partitioned into $N_i$ regions. For region $j \in \{1, \ldots, N_i\}$ in city $i$, the observed outcome, binary treatment, and $p$ proxy variables are denoted by $Y_{ij} \in \mathcal{R}$, $A_{ij} \in \{0,1\}$, and $\bW_{ij} = (W_{ij1}, \ldots, W_{ijp})^\top \in \mathcal{R}^p$. While we present the model with a binary treatment, it can be readily extended to accommodate multi-level treatments (see Section \ref{s:app}). We posit two latent processes to capture confounding. The first is a non-spatial latent confounder process $\bU_{ij} = (U_{ij1}, \ldots, U_{ijq})^\top \in\mathcal{R}^q$, which accounts for unobserved factors influencing both treatment and outcome. The second is a spatial process $Z_{ij} \in \mathcal{R}$, which explains dependence for nearby regions. Consider the model 
\begin{eqnarray}
 Y_{ij} &=& \alpha_y + \theta A_{ij} + \bbb\alpha_{yu}^\top\bU_{ij} + \alpha_{yz} Z_{ij} + \epsilon_{y,ij} \label{eq:y}, \\
A_{ij} &=& \text{I}(\alpha_a + \bbb\alpha_{au}^\top\bU_{ij} + \alpha_{az} Z_{ij} + \epsilon_{a,ij}>0), \label{eq:a} \\
 \bW_{ij} &=& \bbb\alpha_w + \bbb\alpha_{wu}\bU_{ij} + \bbb\epsilon_{w,ij},  \label{eq:w}
\end{eqnarray}
where the error terms $\epsilon_{y,ij}$, $\epsilon_{a,ij}$, and $\bbb\epsilon_{w,ij}$ are independent and identically distributed with mean zero and variances $\sigma_y^2$, $\sigma_a^2$, and $\bbb\Sigma_w = \text{diag}(\sigma_{w_1}^2, \ldots, \sigma_{w_p}^2)$, respectively. Vectors $\bbb\alpha_{au}$ and $\bbb\alpha_{yu}$, both of length $q$, represent the coefficients for the relationships between the latent confounder $\bU_{ij}$ and the treatment $A_{ij}$, and between $\bU_{ij}$ and the outcome $Y_{ij}$, respectively. The matrix $\bbb\alpha_{wu}$ is of dimension $p \times q$, representing the coefficients for the relationships between the proxy variables $\bW_{ij}$ and the latent confounders $\bU_{ij}$. Scalars $\alpha_{az}$ and $\alpha_{yz}$ represent the coefficients between the spatial process $Z_{ij}$ and the treatment $A_{ij}$, and between $Z_{ij}$ and the outcome $Y_{ij}$, respectively. The scalar $\theta$ is the treatment effect we aim to identify and estimate. Intercept terms include the vector $\bbb\alpha_{w}$ and scalars $\alpha_{a}$ and $\alpha_{y}$.

With this design, we acknowledge the presence of latent confounders $\bU_{ij}$ that can be captured through proxy variables $\bW_{ij}$, and account for potential spatial confounding through the shared spatial process $Z_{ij}$. The latent confounder process variables $\bU_{ij}$ are as 
\begin{equation} 
\bU_{ij} \sim \mathcal{D}_u(\bbb\mu_u, \bbb\Sigma_u), \quad \bbb\Sigma_u = \text{diag}(\sigma_{u_1}^2, \ldots, \sigma_{u_q}^2), \label{eq:u} 
\end{equation}
where $\mathcal{D}_u(\cdot)$ means arbitrary distribution with mean and finite variance; without loss of generality, the mean $\bbb\mu_u$ can be set zero. The spatial latent variables are modeled using splines
\begin{equation} 
% Z_{ij} = \sum_{l=1}^{L_{i}} \lambda_{il} B_{il}(\bs_{ij}), \quad 
Z_{ij} = \sum_{l=1}^{L_{i}} \lambda_{il} B_{ijl}, \quad 
\lambda_{il} \sim \mathcal{D}_\lambda(0, \sigma_z^2),
\label{eq:z} \end{equation}
where $\mathcal{D}_\lambda(\cdot)$ means arbitrary distribution with mean and finite variance; $B_{ijl}$ denotes the $l$-th spline basis function integrated over regions $j$ in city $i$ (see Section \ref{s:comp}), and $\lambda_{il}$ are the corresponding coefficients.

\section{Theoretical properties}\label{s:theory}
% paper that guides my proof: Sensitivity analysis with multiple treatments and multiple outcomes with applications to air pollution mixtures

We follow the potential outcomes framework \citep{rubin1976inference} and denote binary treatment as $A$ and outcome as $Y$, then the potential outcomes given a treatment is $Y(A)$. We are interested in the average treatment effect ATE $=\mathbb{E}(\sum_{i=1}^{M}\sum_{j=1}^{N_j}\big\{Y_{ij}(1)-Y_{ij}(0)\big\} / (M \sum_{i=1}^{M} N_i))$. With appropriate assumptions, we show that $\theta$ in Equation (\ref{eq:y}) is the ATE and we can directly use $\hat\theta$ as an ATE estimator. 

\begin{assumption} [SUTVA; Stable Unit Treatment Value Assumption]
\label{sutva}
(1) the potential outcomes for any unit does not vary with the treatment assigned to other units; 
(2) for each unit, there are no different versions of each treatment level that lead to different potential outcomes.
\end{assumption}

\begin{assumption} [Latent Ignorability] %No unmeasured confounding
\label{nuc}
$A_{ij} \indep Y_{ij}(a) | \bU_{ij}, Z_{ij}$ for any $a$. In other words, $\bU_{ij}$ and $Z_{ij}$ account for all confounders influencing treatment and outcome. 
\end{assumption}

\begin{assumption} [Latent Positivity] 
\label{positivity}
$P(A_{ij}=a | \bU_{ij}, Z_{ij}) > 0$ for any $a \in \{0, 1\}$. That is, every unit has a non-zero probability of being assigned any treatment value.
\end{assumption}

\begin{assumption} [Structural Causal Model]
\label{sem} 
The data-generating process is as specified in Equations (\ref{eq:y}) - (\ref{eq:z}).  
%The linear relationships are assumed to hold, though the normality assumptions for the error terms and latent variables, as well as 
The indicator function $I(\cdot)$ in Equation (\ref{eq:a}) can be relaxed. In addition, we assume the two latent process are independent $Z_{ij}\indep \bU_{ij}$. 
\end{assumption}
\begin{assumption}[Sufficient Condition for Factor Model] 
Let $\bbb\Lambda = \bbb\alpha_{wu}\bbb\Sigma_{u|a}^{1/2}$, where $\bbb\Sigma_{u|a}$ represents the conditional variance of $\bU_{ij}$ given treatment $A_{ij} = a$. If any row of $\bbb\Lambda$ is deleted, there remain two disjoint submatrices of rank $q$.
\label{assum_sufficient}
\end{assumption}
%\cite{kang2023sensitivity}

Our results also apply to continuous treatments, with modifications to Assumptions \ref{positivity} and \ref{assum_sufficient}. These continuous counterparts are:

ASSUMPTION 3': \textit{$P(A_{ij}=a | \bU_{ij}, Z_{ij}) > 0$ for any $a \in \mathcal{R}$.}

ASSUMPTION 5': \textit{Let $\bbb\Lambda = \bbb\alpha_{wu}\bbb\Sigma_u^{1/2}$. If any row of $\bbb\Lambda$ is deleted, there remain two disjoint submatrices of rank $q$.} 
% (5 and 5') These conditions can also be expressed as: if any row of $\bbb\alpha_{wu}$ is deleted, there remain two disjoint submatrices of rank $q$; and assume $\bbb\Sigma_u$ is diagonal and positive definite.

\begin{theorem}
Under Assumptions (\ref{sutva}) - (\ref{assum_sufficient}), or by replacing Assumptions (3) and (5) with their continuous counterparts (3') and (5'), the causal effect $\theta$ is identifiable.  
\end{theorem}

We discuss identifiability of model parameters in two scenarios, when the treatment is continuous and binary, respectively. When the treatment is binary, we obtain (derivations in Web Appendix B) 
\begin{align}
& \text{Cov}(\bW_{ij}|A_{ij}) =\bbb\alpha_{wu} \bbb\Sigma_{u|a}\bbb\alpha_{wu}^\top + \bbb\Sigma_w, \label{eq:bi_ww} \\ 
& \text{Cov}(\bW_{ij},Y_{ij}|A_{ij}) = \bbb\alpha_{wu}\bbb\Sigma_{u|a} \bbb\alpha_{yu},
\label{eq:bi_wy} \\ 
& E(\bW_{ij}|A_{ij})=\alpha_a + \bbb\alpha_{wu}E(\bU_{ij}|A_{ij}), \label{eq:bi_ew} \\ 
& E(Y_{ij}|A_{ij}) = \theta A_{ij} + \bbb\alpha_{yu} E(\bU_{ij}|A_{ij}). \label{eq:bi_ey}
\end{align}
In Equation (\ref{eq:bi_ww}), let $\bbb\Lambda = %(I-\Gamma)^{-1}
\bbb\alpha_{wu}\bbb\Sigma_{u|a}^{1/2}$ of shape $p\times q$. We add Assumption \ref{assum_sufficient}, which is a strong and sufficient condition and implies $p \geq 2q + 1$. With Assumption \ref{assum_sufficient} and by applying Lemma 5.1 and Theorem 5.1 of \cite{anderson1956statistical},  then $\bbb\Lambda$ is identified up to rotations from the right under certain sufficient conditions. Specifically, $%(I-\Gamma)^{-1}
\bbb\alpha_{wu}\bbb\Sigma_{u|a}^{1/2}$ is identified up to multiplication on the right by an orthogonal matrix $\bR_1$, so any admissible value for $\bbb\Lambda$ can be written as $\bbb\Lambda^* = \bbb\Lambda \bR_1$ with an arbitrary orthogonal matrix $\bR_1$ of shape $q \times q$ (\cite{miao2023identifying}, \cite{kang2023sensitivity}). 

Plugging $\bbb\Lambda$ into Equation (\ref{eq:bi_wy}), it becomes a linear system with $p$ equations and $q$ unknowns. $\bbb\Sigma_{u|a}^{1/2}\bbb\alpha_{yu}$ is identified up to multiplication on the left by $\bR_1^\top$. Similarly, plugging $\bbb\Lambda$ into Equation (\ref{eq:bi_ew}), then $E(\bU_{ij}|A_{ij})$ is identified up to multiplication on the left by $\bR_1^\top\bbb\Sigma^{-1/2}_{u|a}$. 
Finally in Equation (\ref{eq:bi_ey}), $\bbb\alpha_{yu}^\top E(\bU_{ij}|A_{ij})$ has been identified since it can be expressed by $(\bbb\alpha_{yu}^\top\bbb\Sigma_{u|a}^{1/2}\bR_1)(\bR_1^\top\bbb\Sigma_{u|a}^{-1/2}E(\bU_{ij}|A_{ij}))$, which are two components that have been identified. Consequently, the causal effect $\theta$ in Equation (\ref{eq:bi_ey}) can be uniquely identified. When spatial confounding $Z$ exists, we approximate it by B-splines (see details in Section \ref{s:comp1}). % we follow \cite{guan2023spectral} for identification. The key assumption is that confounding present at global scales dissipates at local scales. 
Equation (\ref{eq:bi_wy}) and (\ref{eq:bi_ey}) will be updated as below, while the method proof and conclusion remain the same. 
\begin{align}
& \text{Cov}(\bW_{ij},Y_{ij}|A_{ij}) = \bbb\alpha_{wu}\bbb\Sigma_{u|a} \bbb\alpha_{yu} + \text{Cov}(\bW_{ij},Z_{ij}|A_{ij}) \alpha_{yz}, \nonumber
 \\ 
& E(Y_{ij}|A_{ij}) = \theta A_{ij} + \bbb\alpha_{yu} E(\bU_{ij}|A_{ij}) + \alpha_{yz} E(Z_{ij}|A_{ij}). \nonumber
\end{align}

When the treatment is continuous, we obtain (derivations in Web Appendix B) 
\begin{align}
& \text{cov}(\bW_{ij})
= \bbb\alpha_{wu}\bbb\Sigma_u\bbb\alpha_{wu}^\top+\bbb\Sigma_w,
\label{eq:ww} \\ 
& \text{cov}(\bW_{ij}, Y_{ij}) 
= \text{cov}(\bW_{ij}, A_{ij}) \theta + \bbb\alpha_{wu} \bbb\Sigma_u \bbb\alpha_{yu}.
\label{eq:wy}
\end{align} 
In Equation (\ref{eq:ww}), let $\bbb\Lambda = %(I-\Gamma)^{-1}
\bbb\alpha_{wu}\bbb\Sigma_u^{1/2}$ of shape $p\times q$. We add Assumption 5', then by applying Lemma 5.1 and Theorem 5.1 of \cite{anderson1956statistical},  $\bbb\Lambda$ is identified up to multiplication on the right by orthogonal matrix $\bR$. In Equation (\ref{eq:wy}), denote $\bbb\Xi = \bbb\Sigma_u^{1/2} \bbb\alpha_{yu}$, resulting in a linear system with $p$ equations and $(1+q)$ unknowns of $\theta$ and $\bbb\Xi$. With the same Lemma 5.1 and Theorem 5.1 by \cite{anderson1956statistical}, $p \geq 2q + 1$ holds, therefore, these equations are over-determined and can be solved uniquely for $\theta$ using $\text{cov}(\bW_{ij}, Y_{ij}) = \text{cov}(\bW_{ij}, A_{ij}) \theta + \bbb\Lambda^* \bbb\Xi$. 

\section{Computational details}\label{s:comp}

\subsection{Approximation of spatial confounders}\label{s:comp1} 

We assume an independent spatial latent process for each city,
and approximate spatial confounding $Z_{ij}$ by a linear combination of B-splines, $Z_{ij} = \sum_{l=1}^{L_i} \lambda_{il} B_{ijl}$, where $B_{ijl}$ is the $l$-th pre-computed spline basis function for region $j$ of city $i$ and $\lambda_{il}$ is the corresponding coefficient. The number of basis function is taken to be $L_i = \lfloor r N_i \rfloor$ where $r$ is the ratio of the number of basis functions to the number of regions in a city. The ratio $r$ is selected by minimizing  Watanabe-Akaike Information Criterion (WAIC) of the outcome model \citep{gelman2014understanding}.

% WAIC is a Bayesian measure of model fit that balances goodness of fit with model complexity, making it ideal for comparing models of varying complexity in our study. Specifically, we measure WAIC at the outcome model, where $Y$ is generated given $\{U_i\}_{i=1}^q, Z, A$. We compute the same statistics as above for these estimates. 

To create spline basis functions for city $i$, we define the minimum bounding rectangle $\mathcal{G}_i$ that encompasses city $i$, and place a 100-by-100 grid of points within $\mathcal{G}_i$, denoted by coordinates $\bbb s_{k}$ for $k = \{1,2,...,10000\}$. Then we construct 2D cubic splines using this coordinates, denoted $b_{il}(\bbb s)$ for $l = \{1,...,L_i\}$. Finally, within polygon $A_{ij}$, we integrate these splines over locations $\bbb s_k \in A_{ij}$, resulting in the spline basis $B_{ijl} = \sum_k b_{il}(\bbb s_{k})\textbf{1}(\bbb s_k \in A_{ij}) / \sum_k \textbf{1}(\bbb s_k \in A_{ij})$. % B_{ijl} = \int_{\bbb s_k \in A_{ij}} b_{il}(\bbb s_{k})d(\bbb s)$.

\subsection{Bayesian Markov Chain Monte Carlo}\label{s:comp2}
We use Bayesian methods to incorporate uncertainties and address the inherent challenges in the complex data structure, including spatial and non-spatial latent variables, and zero-inflated proxies. We use a Markov chain Monte Carlo (MCMC) approach to sample from the joint posterior distribution of our model. Standard techniques for MCMC are employed, and uninformative priors are used when necessary. 

For simulation studies, we run single chain MCMC with 50,000 burn-in iterations and 50,000 iterations post burn-in, with a thinning factor of 10. For real data analysis, we run single chain MCMC with 150,000 burn-in iterations and 150,000 more post burn-in, with a thinning factor of 10. Convergence is monitored using trace plots. Further details are provided in Web Appendix C.

\section{Simulation study}\label{s:sim}
The objectives of the simulation study are to evaluate the performance of our model in terms of estimation and inference. We conduct the simulation using two settings: (1) creating data with simple grid geometry and predetermined parameters, and (2) creating data that closely resemble the redlining data. % For data generation, we use a true spline ratio of either 40\% or 60\% in different settings. 
For each parameter setting, we randomly generate 100 datasets. % More details about data generation is in section \ref{s:sim:gen}.

\subsection{Data generation}\label{s:sim:gen}
% keep this note %
% Case (1) simu1
% Case (2) simu9
% Case (3) simu0
% Case (4) simu2
% Case (5) simu7
% Case (6) simu8 
% Case (7) simu10 
    % Case (7) simu3 drop 
% Case (a) and (b) simured 
% Case (8) simu4
% Case (9) simu5
% Case (10) simu6

To create data with simple grid geometry and predetermined parameters, we generate 490 regions in 10 cities, consisting of 7-by-7 grid regions in each city. The data-generation process is defined by
\begin{align}
&U_{ij} \sim \mathcal{N}(0, 1), \quad Z_{ij} = \sum_{l=1}^{L_i} \lambda_{il} B_{ijl}, \quad  \lambda_{il} \sim \mathcal{N}(0, 1), \nonumber\\ 
&W_{ij1} \sim \mathcal{N}(\alpha_{w_1u}U_{ij}, \sigma^2_{w_1}), \quad W_{ij2} \sim \mathcal{N}(\alpha_{w_2u}U_{ij}, \sigma^2_{w_2}),  \nonumber\\ 
&\tilde W_{ij3} \sim \mathcal{N}(\alpha_{w_3u}U_{ij}, \sigma^2_{w_3}), \quad W_{ij3} = \max(\tilde W_{ij3}, \text{quantile}_{.05}(\tilde W_3)), \nonumber\\ 
&\text{logit}(P(A_{ij} = 1)) = \alpha_{a} + \alpha_{au} U_{ij} + \alpha_{az} Z_{ij}, \nonumber\\
&Y_{ij} \sim \mathcal{N}(\theta A_{ij} + \alpha_{yu} U_{ij} + \alpha_{yz} Z_{ij}, \sigma^2_y). \nonumber
\end{align}

We consider six cases with correctly specified model. The first is a base case with 
(1) $\theta=0.2$, $\alpha_{w_1u}=2$, $\alpha_{w_2u}=-0.5$, $\alpha_{w_3u}=-1$, $\sigma^2_{w_1}=\sigma^2_{w_2}=\sigma^2_{w_3}=1$, $r=40\%$, $\alpha_{a}=-1$, $\alpha_{au}=-1$, $\alpha_{az}=1$, $\alpha_{yu}=-0.2$, $\alpha_{yz}=-1$, $\sigma^2_y=0.25$. The others cases modify the base case as follows: (2) Stronger proxy: $\sigma^2_{w_1}=\sigma^2_{w_2}=\sigma^2_{w_3}=0.5$,
(3) Noisier outcome: $\sigma^2_y=1$, and (4) Rougher spatial confounding: $r=60\%$. 
(5) Weaker proxy: $\sigma^2_{w_1}=\sigma^2_{w_2}=\sigma^2_{w_3}=2$, 
(6) Stronger confounder $\bU$: $\alpha_{au} = -2, \alpha_{yu} = -2$. 
% (7) $\bU$ as a spatial process. 

We further have four cases with incorrectly specified model revised upon case (1): 
(7) Misspecified model on $\mathbb{E}(A |U)$ and $\mathbb{E}(Y |U)$: $A$ and $Y$ depends on $W_1, W_2, W_3$ directly, generate data with $\text{logit}(P(A_{ij} = 1)) = \alpha_{a} + \alpha_{au} (-W_{ij1}+W_{ij2}+W_{ij3}) + \alpha_{az} Z_{ij}$, $Y_{ij} \sim \mathcal{N}(\theta A_{ij} + \alpha_{yu} (-W_{ij1}+W_{ij2}+W_{ij3}) + 0.2U_{ij}Z_{ij} + \alpha_{yz} Z_{ij}, \sigma^2_y)$.
(8) Misspecified model on $\mathbb{E}(Y |U, Z)$: generate data with $Y_{ij} \sim \mathcal{N}(\theta A_{ij} + \alpha_{yu} U_{ij} + \alpha_{yz} Z_{ij} + 0.2U_{ij}Z_{ij}, \sigma^2_y)$,
(9) Misspecified model on $\mathbb{E}(Y |U)$: generate data with $Y_{ij} \sim \mathcal{N}(\theta A_{ij} + \alpha_{yu} U_{ij}^2 + \alpha_{yz} Z_{ij}, \sigma^2_y)$, 
(10) Misspecified model on $\mathbb{E}(Y |U)$: generate data with $Y_{ij} \sim \mathcal{N}(\theta A_{ij} + \alpha_{yu} log(|U_{ij}|) + \alpha_{yz} Z_{ij}, \sigma^2_y)$. 
In all cases, $W_{ij3}$ is generated to have the lowest 5\% percent values as zero, to model the zero-inflated percentage of Black Population in the real data. 

To generate data that closely resemble the redlining data, we use the geometry of the redlining data. For each dataset, we take a random subset cities from the real data such that there are at least 500 regions in a dataset. We define true parameters as the posterior parameter estimates from real Redlining data analysis (using $r=60\%$). There are two cases: (a) outcome of NO$_2$, (b) outcome of PM$_{2.5}$. 

\subsection{Competing methods and metrics}\label{s:sim:methods}

We compare our method with two alternatives: (1) No adjustment for latent SES, removing latent process $\bU_{ij}$ and proxy $\bW_{ij}$ from the model; (2) Outcome Regression with Proxy, removing $\bU_{ij}$ and moving proxy $\bW_{ij}$ into the outcome regression as covariates. Both methods contains the spatial latent process $Z_{ij}$. We run models using spline ratios $r=\{0\%, 20\%, 40\%, 60\%, 80\%\}$ to explore a broad range of complexities in the spatial latent processes, and select the best model based on WAIC. 

We denote the true effect as $\theta^*$. For each method, we denote the effect estimates as $\hat{\theta}_t$ for simulation data set $t$, $t \in \{1,...,100\}$, and $(L_t, U_t)$ as the corresponding credible intervals. We compute the following statistics: absolute bias $= 100^{-1} |\sum_{t=1}^{100}(\hat{\theta}_t - \theta^*)|$, mean squared error (MSE) $= 100^{-1}\sum_{t=1}^{100}(\hat{\theta}_t - \theta^*)^2$, and coverage probability $=100^{-1}  \sum_{t=1}^{100} \mathbb{I} (\theta^* \in [L_t, U_t])$.

\subsection{Results}\label{s:sim:results}

The simulation results are shown in Table \ref{t:sim1}, including the absolute bias (A.B.), mean squared error (MSE), coverage probability (\%, C.P.), WAIC optimized over $r$, and the spline ratio (S.R.), each averaged over 100 simulations. Cases (1) - (4) use fully synthetic data, and cases (a) and (b) closely mimics the redlining data. In cases (1)-(4), our method obtains a satisfying coverage probability and negligible absolute bias and MSE. By minimizing WAIC, on average our method selects a spline ratio $r$ that is only slightly higher than the true ratio. In cases (a) and (b), the coverage probability is $90\%$ for NO$_2$ and $91\%$ for PM$_{2.5}$. Comparing across methods, our approach consistently produces coverage probabilities closest to the nominal 95\% level, along with the lowest absolute bias and MSE. In summary, our method outperforms the alternatives across all evaluated metrics and simulation scenarios in Table \ref{t:sim1}. 

For additional robustness checks, we provide further simulations in Web Table S2. These simulations explore scenarios with weaker proxies, stronger confounding, and model misspecification. We observe minor performance drops in cases of weak proxies and strong confounding, which could be mitigated by adjusting the spline ratio. However, the performance drops dramatically upon model misspecification, that is, when the structural assumption (\ref{assum_sufficient}) is violated. 

\begin{table}
\centering
\caption{Simulation results by cases: (1) base case, (2) stronger proxy, (3) noisier outcome, (4) rougher spatial confounding, (a) use posterior parameters and outcome is NO$_2$, (b) use posterior parameters and outcome is PM$_{2.5}$. In case (1), (2), (3), the true spline ratio is 40\%; in case (4), (a), (b), the true spline ratio is 60\%. The columns display the average absolute bias (A.B.) with standard deviation, mean square error (MSE) with standard deviation, coverage probability (C.P.), Watanabe-Akaike Information Criterion (WAIC), and the selected spline ratio (S.R.).}
\begin{tabular}{llrrrrr}
\hline\hline
Case & Method & A.B. & MSE & C.P. & WAIC & S.R. \\  
\hline
(1) & Latent Adjustment  & 0.003 (0.106) & 0.011 (0.015) & 95 & 1083 & 54 \\ 
& Outcome Regr with Proxy & 0.245 (0.088) & 0.068 (0.043) & 18 & 1115 & 44 \\ 
& No Adjustment & 1.197 (0.106) & 1.445 (0.257) & 0 & 1460 & 42 \\ 
\hline
(2) & Latent Adjustment & 0.003 (0.086) & 0.007 (0.010) & 94 & 1030 & 49 \\ 
& Outcome Regr with Proxy & 0.135 (0.078)& 0.024 (0.022) & 55 & 1038 & 44 \\ 
& No Adjustment & 1.198 (0.102) & 1.446 (0.246) & 0 & 1460 & 42 \\ 
\hline
(3) & Latent Adjustment & 0.010 (0.159) & 0.025 (0.040) & 95 & 1590 & 43 \\ 
& Outcome Regr with Proxy & 0.228 (0.125) & 0.067 (0.060) & 54 & 1580 & 37 \\ 
& No Adjustment & 1.170 (0.143) & 1.388 (0.341) & 0 & 1736 & 40 \\ 
\hline
(4) & Latent Adjustment & 0.012 (0.107) & 0.011 (0.014) & 95 & 110 & 66 \\ 
& Outcome Regr with Proxy & 0.243 (0.090) & 0.067 (0.041) & 19 & 1144 & 58 \\ 
& No Adjustment & 1.198 (0.103) & 1.445 (0.249) & 0 & 1485 & 60 \\ 
\hline
(a) & Latent Adjustment & 0.001 (0.154) & 0.023 (0.036) & 90 & 1874 & 69 \\ 
& Outcome Regr with Proxy & 0.051 (0.215) & 0.052 (0.079) & 78 & 1905 & 68 \\ 
& No Adjustment & 0.640 (0.124) & 0.425 (0.158) & 0 & 1967 & 69 \\ 
\hline
(b) & Latent Adjustment & 0.003 (0.040) & 0.002 (0.002) & 91 & 470 & 69 \\ 
& Outcome Regr with Proxy & 0.004 (0.063) & 0.004 (0.005) & 73 & 498 & 68 \\ 
& No Adjustment & 0.048 (0.032) & 0.003 (0.003) & 67 & 478 & 68 \\
\hline
\end{tabular}
\label{t:sim1}
\end{table}

\section{Redlining policy analysis}\label{s:app}
We apply our method to study the effect of redlining policy on air pollution exposure. To account for socio-economic status in the 1930s, we define three proxy variables: the box-cox transformed unemployment rate, the mean house rent, and the rank-based inverse normal transformed percent of Black population. We define the control group as grade A and the treatment groups as grades B, C, and D, denoting their respective treatment effects as ‘B-A', ‘C-A', and ‘D-A'. The outcomes are PM$_{2.5}$ and NO$_2$ concentrations in 2010, and we fit our model separately for each pollutant. All proxy variables and outcomes are centered by their mean per city before fitting the model. The model in Equation (\ref{eq:y})-(\ref{eq:w}) is extended to have three binary treatment variable by adding a spatial term for each treatment variable. In addition to the constant causal effect model discussed in Section \ref{s:method}, we also develop a random effect model, assuming that causal effects vary by city and are independent and identically distributed. Further details on models, including multiple treatments and random effects, can be found in Web Appendix C. 

\subsection{Results}\label{s:app:results}

\begin{figure}
\centering
\includegraphics[width=6cm]{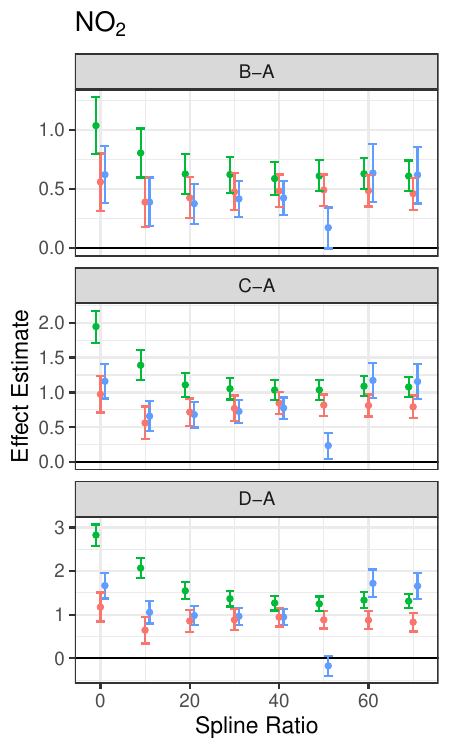}
\includegraphics[width=10cm]{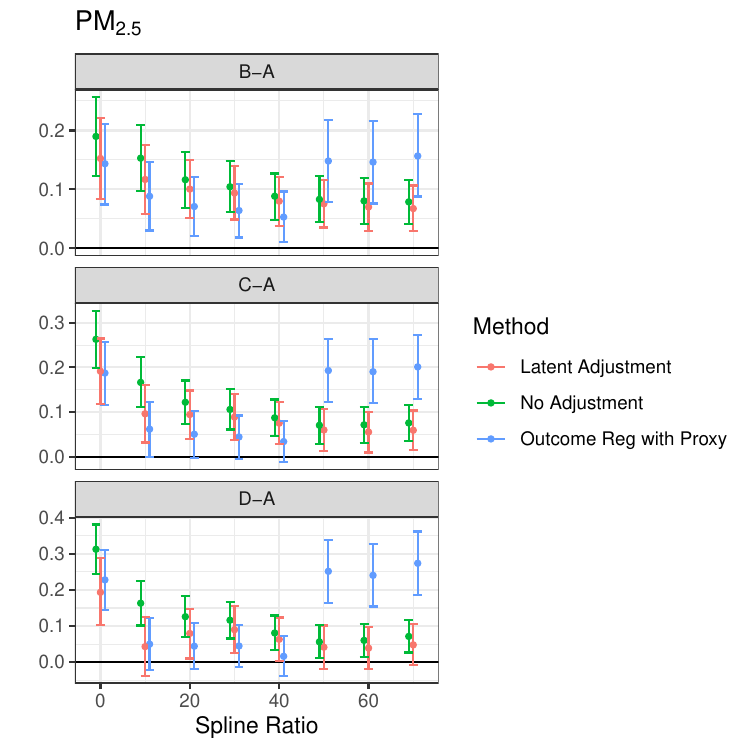}
\caption{Redlining effect estimates and 95\% credible intervals. The left panel shows results for NO$_2$ (ppb) and the right panel for PM$_{2.5}$ ($\mu$g/m$^3$). Estimates are shown for different redlining grades relative to grade A (`B-A', `C-A', `D-A') under varying spline ratios.}
\label{fig:red_estimate}
\end{figure}

Figure \ref{fig:red_estimate} shows the posterior distribution of the long-term effects of redlining policies on air pollution exposure. The estimates stabilize as the spline ratio increases for the Latent Adjustment and No Adjustment methods, indicating robustness across these approaches. However, the Outcome Regression with Proxy method shows greater variability, suggesting potential sensitivity to the spline choice. 
 
For most treatment groups and pollutants, the estimated effects are significant and positive, indicating that redlining had a harmful impact on air quality. Specifically, at $r=60\%$, the estimated effect for NO\(_2\) is 0.48 ppb (95\% CI: 0.35 to 0.62) for `B-A', 0.81 (0.65 to 0.96) for `C-A', and 0.87 (0.67 to 1.08) for `D-A'. The estimated effect for PM\(_{2.5}\) is 0.07 \(\mu g/m^3\) (95\% CI: 0.03 to 0.11) for `B-A', 0.06 (0.01 to 0.10) for `C-A', and 0.04 (-0.02 to 0.10) for `D-A'. This is different from the raw difference between all treatment groups (B, C, D) and control group A, which is 2.48 ppb for NO$_2$ and 0.26 $\mu$g/m$^3$ for PM$_{2.5}$. Overall, the Latent Adjustment method provides stable estimates, while the Outcome Regression with Proxy method exhibits substantial variability. If our model correctly represents the data-generating process, the No-Adjustment method tends to overestimate the effect for NO$_2$. 

Web Figure S4 demonstrates that Latent Adjustment achieves lowest WAIC values compared to the No Adjustment and Outcome Regression with Proxy. There is a rapid decrease in WAIC values until spline ratio $r$ reaches approximately 60\%. Given the stable estimates observed in Figure \ref{fig:red_estimate}, we will highlight results at $r=60\%$ for the remainder of this paper.

\begin{figure}[!htbp]
\centering
\includegraphics[width=8cm]{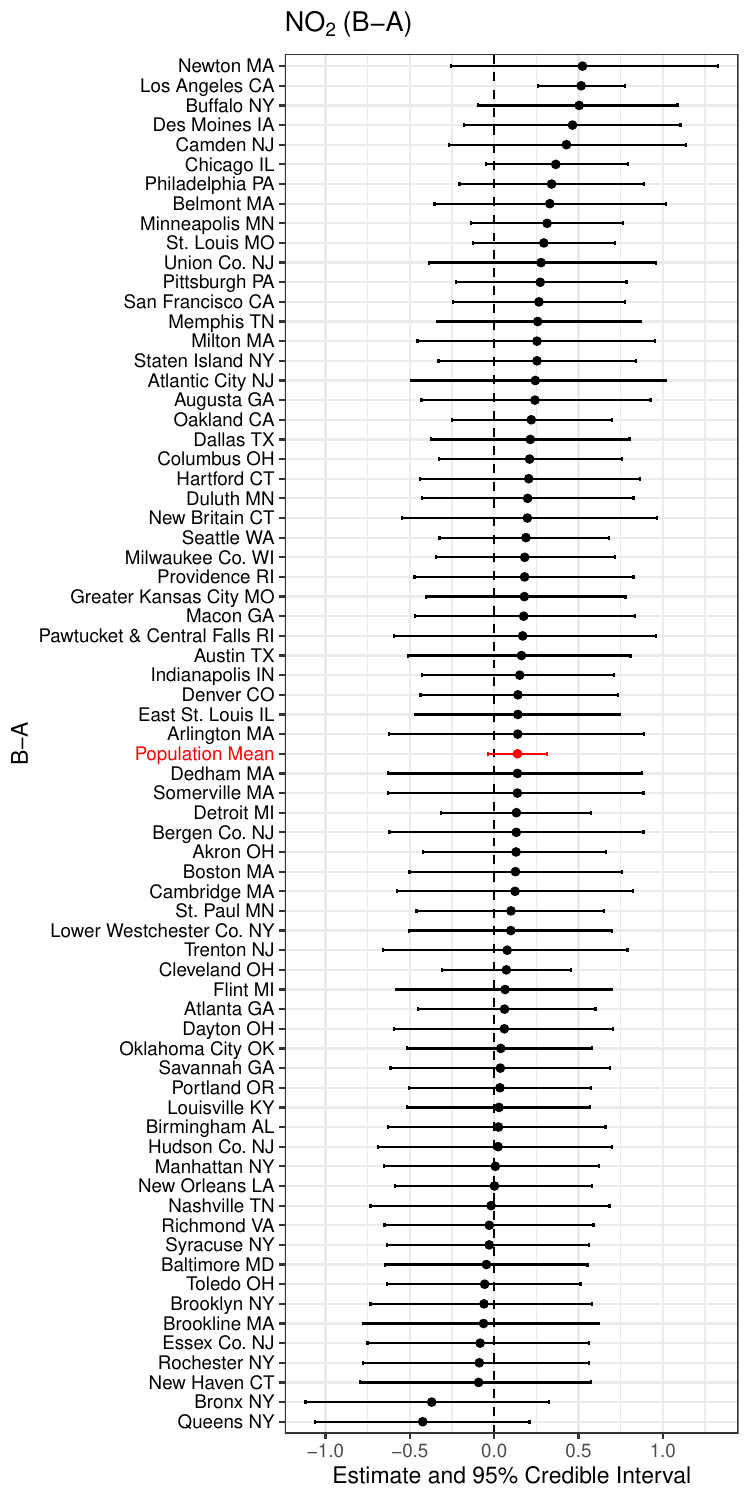}
\includegraphics[width=8cm]{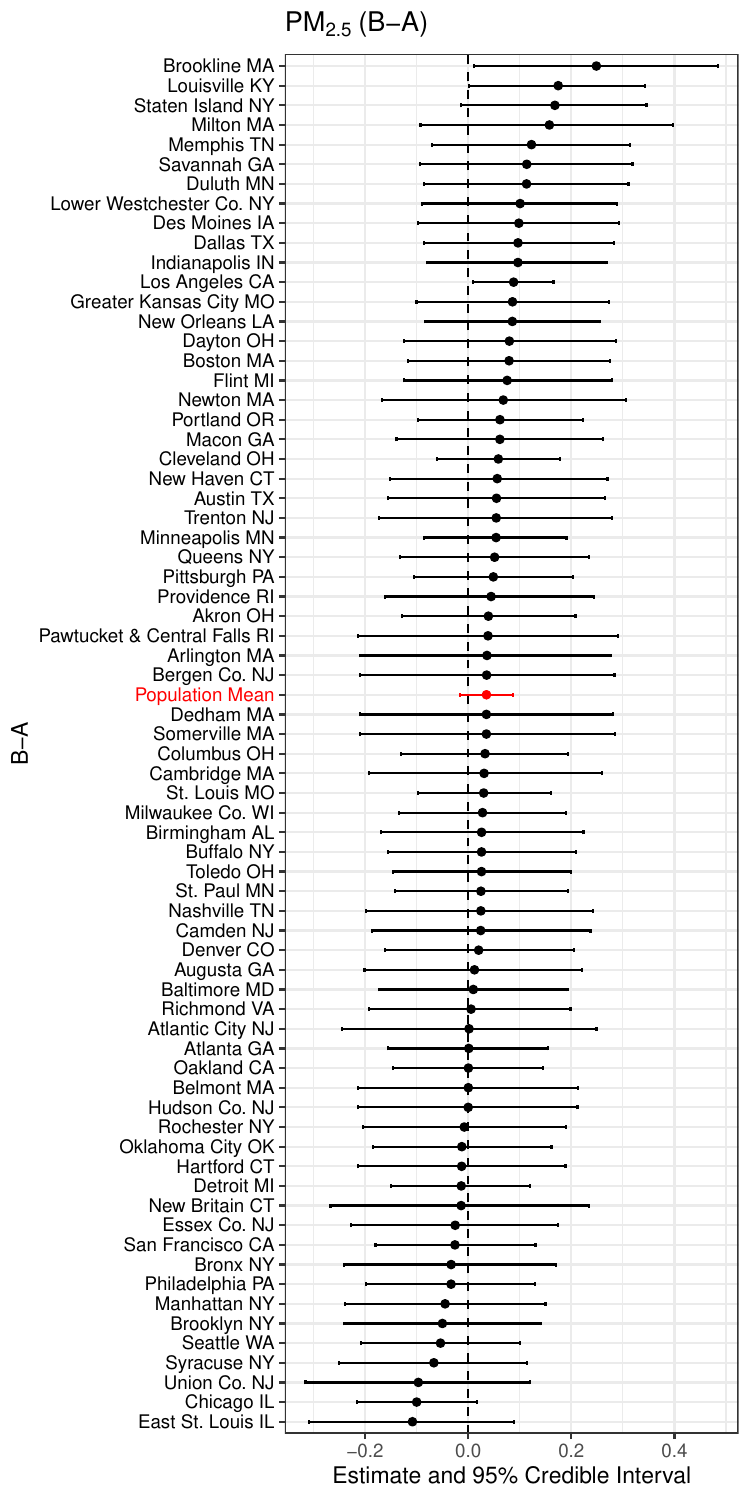}
\caption{Posterior estimates of the long-term effects of `B-A' on NO$_2$ (left) and PM$_{2.5}$ (right) concentrations across 69 cities. Each city is represented by its posterior mean and 95\% credible interval. The population mean is shown in red.} 
\label{fig:red_random_effect_ba}
\end{figure}

\begin{figure}[!htbp]
\centering
\includegraphics[width=8cm]{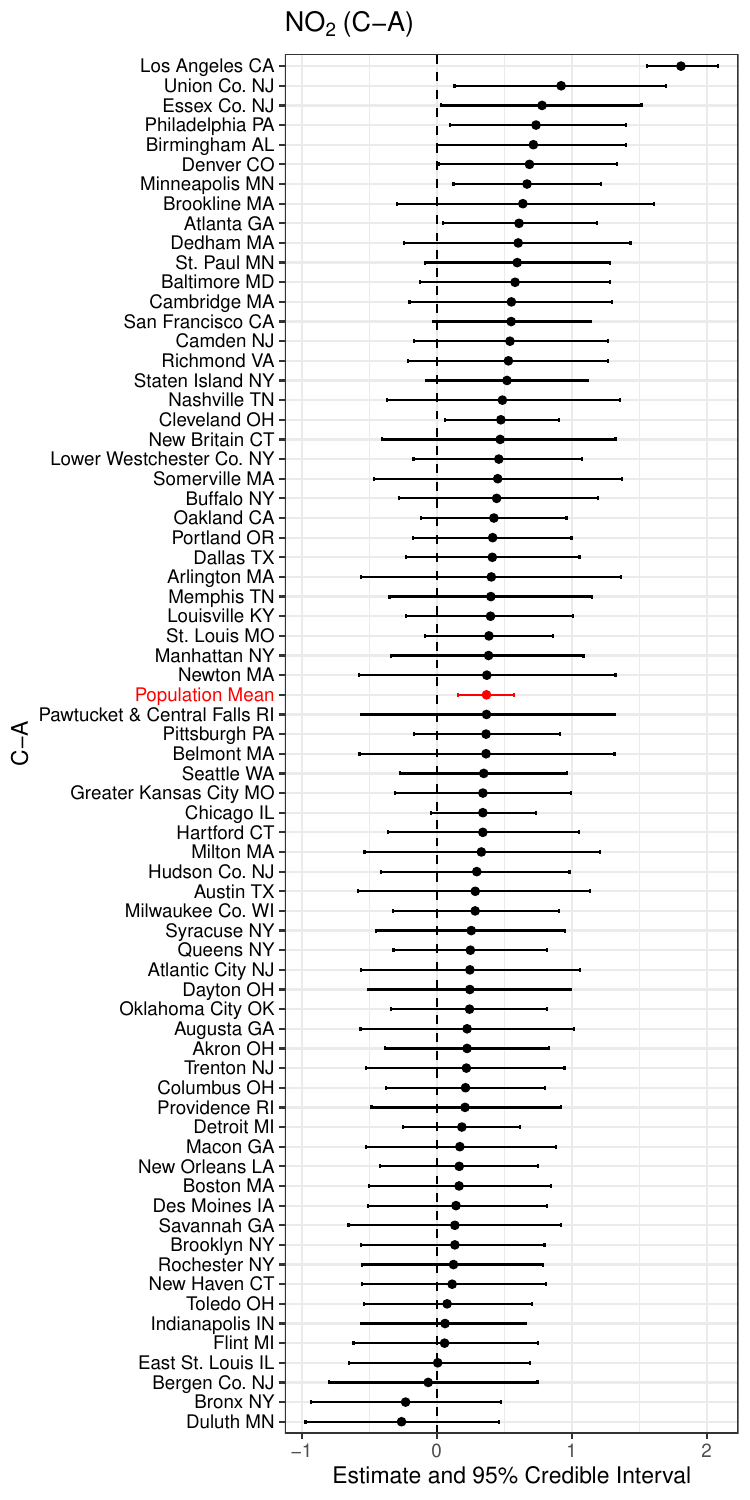}
\includegraphics[width=8cm]{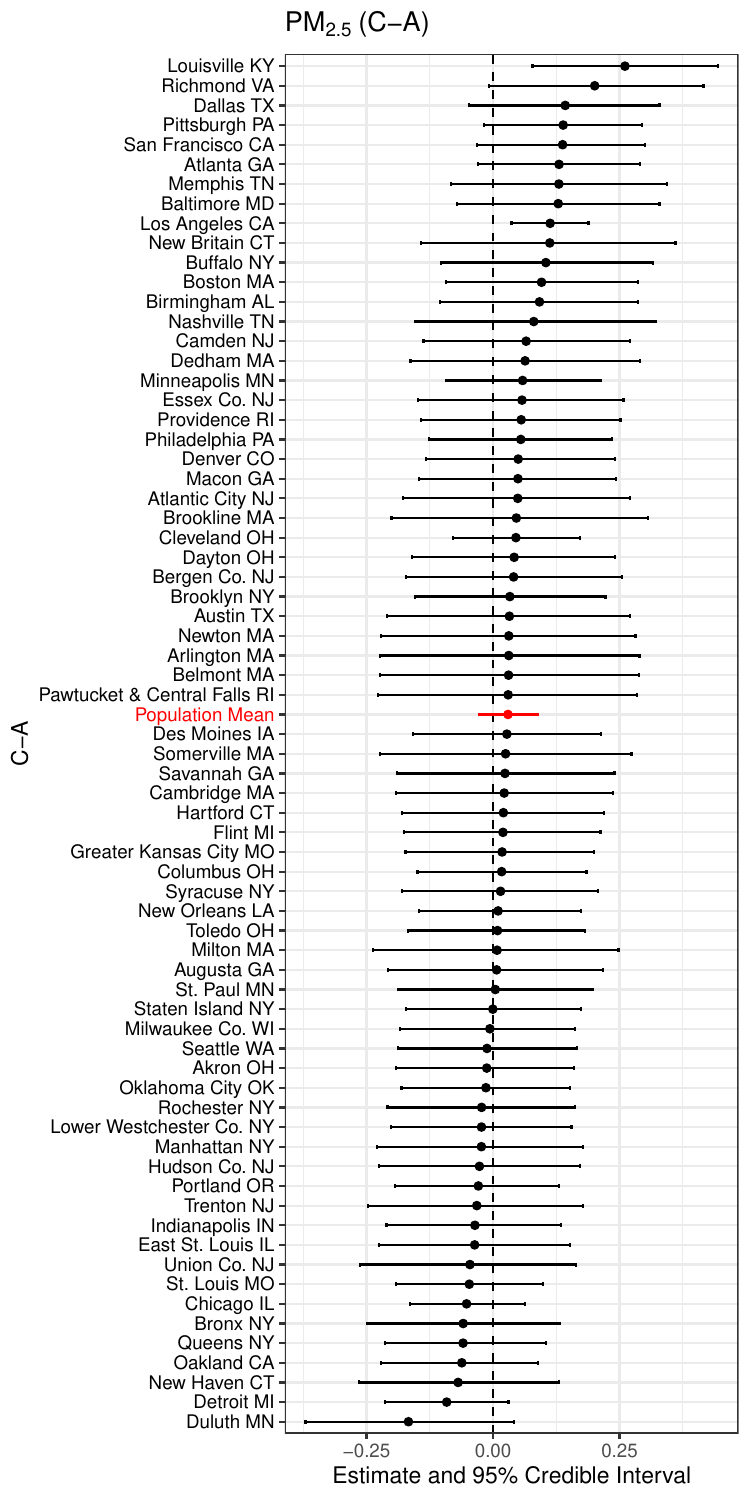}
\caption{Posterior estimates of the long-term effects of `C-A' on NO$_2$ (left) and PM$_{2.5}$ (right) concentrations across 69 cities. Each city is represented by its posterior mean and 95\% credible interval. The population mean is shown in red.} 
\label{fig:red_random_effect_ca}
\end{figure}

\begin{figure}[!htbp]
\centering
\includegraphics[width=8cm]{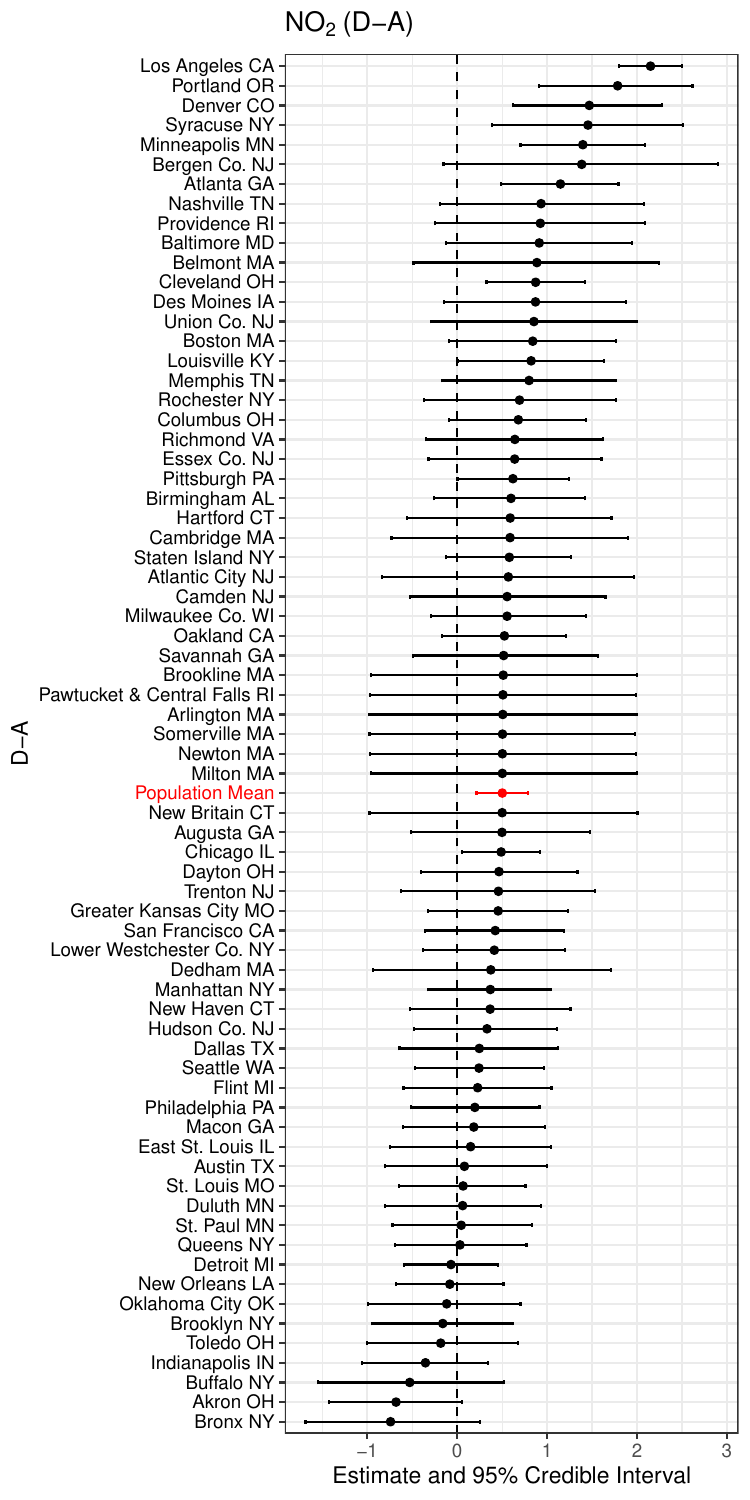}
\includegraphics[width=8cm]{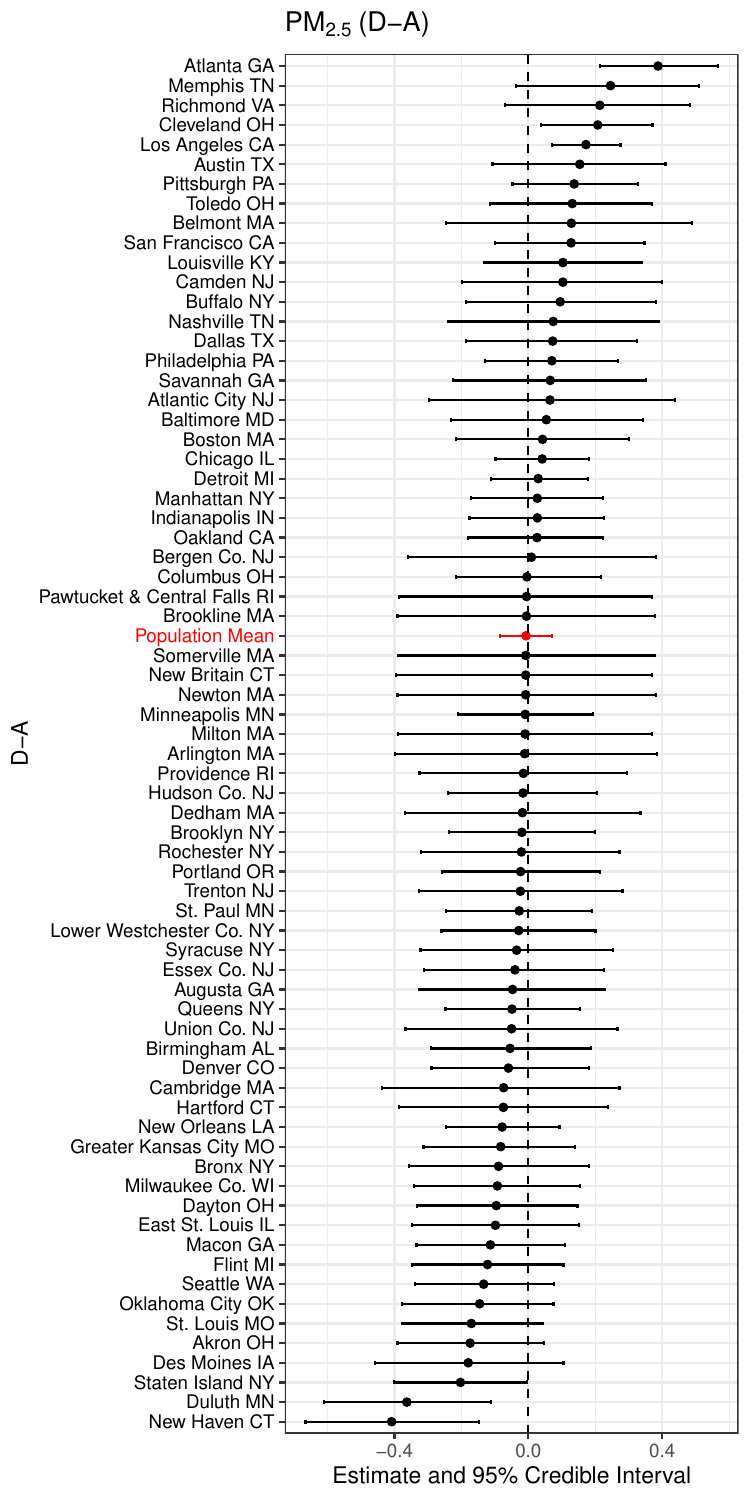}
\caption{Posterior estimates of the long-term effects of `D-A' on NO$_2$ (left) and PM$_{2.5}$ (right) concentrations across 69 cities. Each city is represented by its posterior mean and 95\% credible interval. The population mean is shown in red.} 
\label{fig:red_random_effect_da}
\end{figure}

We apply a random effect model with $r=60\%$, where the results of the constant effects model had stabilized over a wide range of $r$ values. For NO\(_2\), the estimated average treatment effects cross cities are 0.14 ppb (95\% CI: -0.04 to 0.31) for `B-A', 0.37 (0.16 to 0.57) for `C-A', and 0.50 (0.22 to 0.79) for `D-A'. These estimates suggest an increasing impact of redlining on NO\(_2\) exposure across treatment levels, with stronger effects observed for more heavily redlined areas. For PM\(_{2.5}\), the estimated average treatment effects cross cities are 0.04 \(\mu g/m^3\) (95\% CI: -0.02 to 0.09) for `B-A', 0.03 (-0.03 to 0.09) for `C-A', and -0.01 (-0.08 to 0.07) for `D-A'. These results indicate no strong evidence of a redlining effect on PM\(_{2.5}\), with credible intervals spanning zero for all treatment levels. Overall, the random effects model confirms a statistically significant and increasing effect of redlining on NO\(_2\) exposure, while the effects on PM\(_{2.5}\) remain weak and inconclusive.

As shown in Figures \ref{fig:red_random_effect_ba}, \ref{fig:red_random_effect_ca}, and \ref{fig:red_random_effect_da}, several cities exhibit strong evidence of a harmful effect on NO\(_2\) among different treatment groups (`B-A', `C-A', `D-A'), including Los Angeles, CA; Philadelphia, PA; Minneapolis, MN; Denver, CO; Atlanta, GA; Portland, OR; and Cleveland, OH. No cities show evidence of a protective effect for NO\(_2\). For PM\(_{2.5}\), the effects are generally weaker, but some cities still show significant disparities. Los Angeles, CA, and Atlanta, GA exhibit strong evidence of a harmful effect at the `D-A' comparison, while New Haven, CT, and Duluth, MN, show strong evidence of a protective effect. Notably, Los Angeles and Atlanta consistently show the strongest harmful effects for both NO\(_2\) and PM\(_{2.5}\), highlighting the persistent environmental impact of redlining in these cities.  
 
\begin{figure}[!htbp]
\centering
\includegraphics[width=8cm]{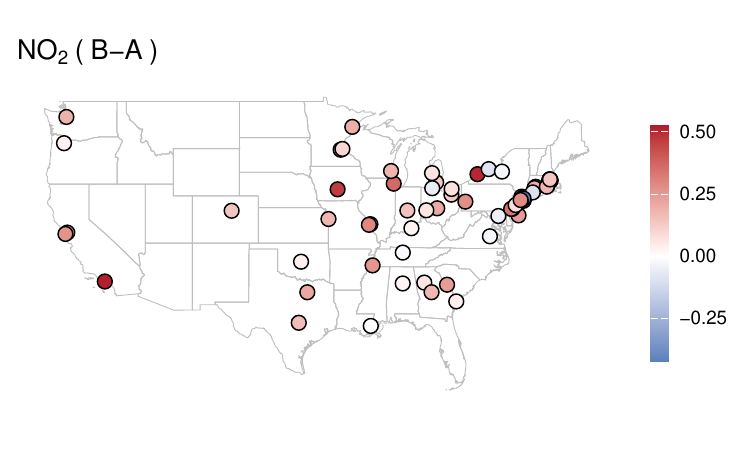}
\includegraphics[width=8cm]{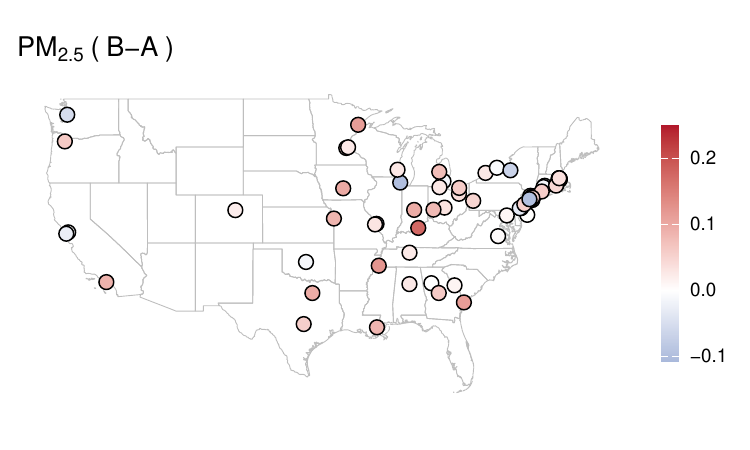}
\includegraphics[width=8cm]{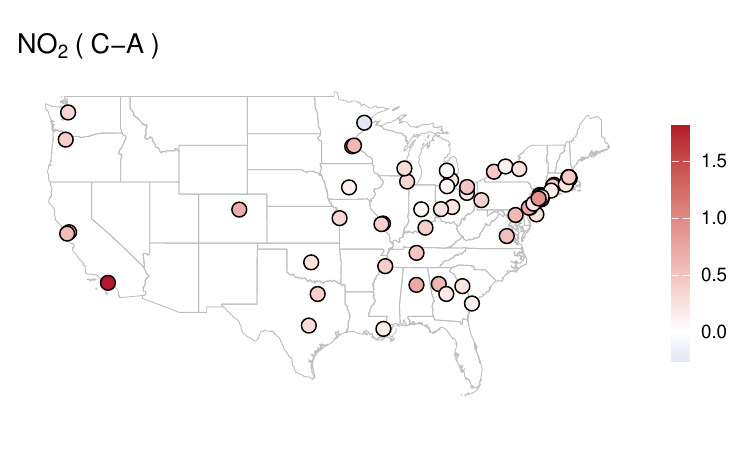}
\includegraphics[width=8cm]{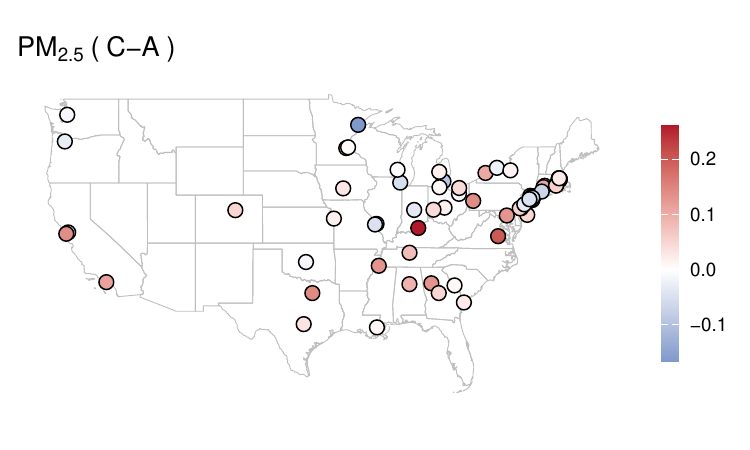}
\includegraphics[width=8cm]{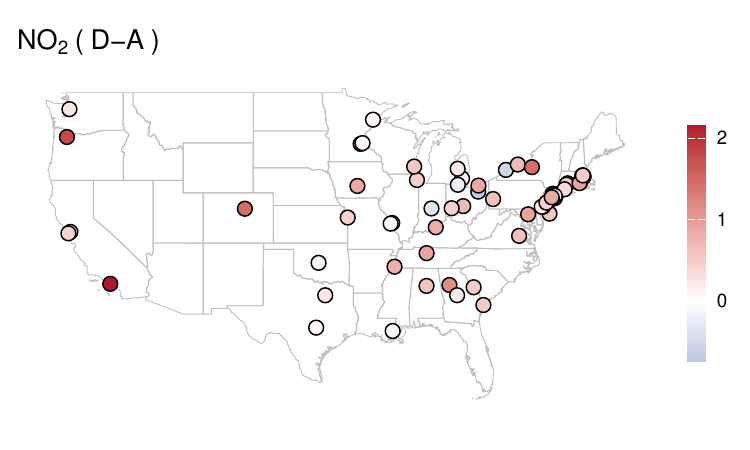}
\includegraphics[width=8cm]{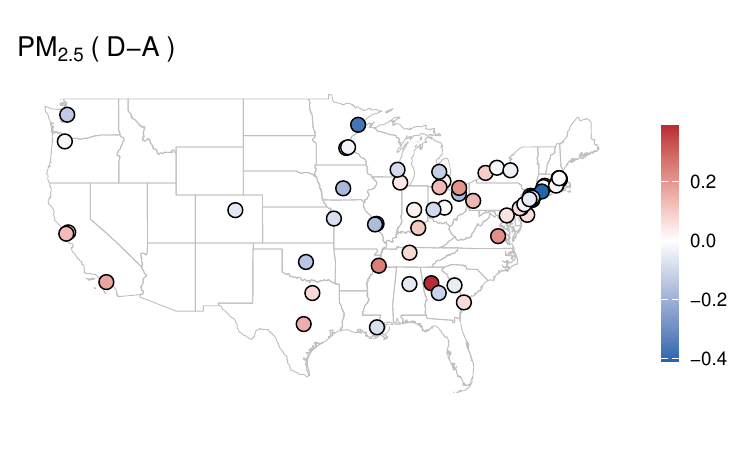}
\caption{Mapping posterior estimates of the random effects across 69 cities. Harmful effects are represented by red and protective effects by blue, regardless of significance.}
\label{fig:map_random_effect}
\end{figure}

The spatial distribution of the long-term effects across 69 cities, as depicted in Figure \ref{fig:map_random_effect}, reveals distinct geographic patterns. For PM$_{2.5}$, the harmful effects, indicated by red dots, are predominantly concentrated along the East Coast and in certain Midwestern and Western cities. Conversely, protective effects, represented by blue dots, are more apparent in central and northern cities. The relationship between these spatial patterns and urban development and population trends warrants further investigation. For NO$_2$, the spatial distribution of harmful effects is much broader, encompassing a wide range of geographic regions. 

In Web Figure S3 and Web Table S3, we confirm that the overlap and positivity assumptions are met, ensuring a solid foundation for causal inference. Web Table S4 demonstrates that we identify the latent $U_{ij}$ representing socio-economic status (SES) in the expected manner. A higher value in latent factor $U_{ij}$ indicates lower socio-economic status: $U_{ij}$ is associated with a higher unemployment rate, lower house rent, higher percentage of Black population, higher probability of being redlined, and higher air pollution concentrations.

\section{Discussion}\label{s:discussion}

We estimate the long-term causal effects of redlining policies (1935-1974) on present-day NO$_2$ and PM$_{2.5}$ air pollution concentrations in 69 cities across 27 U.S. states. We found strong evidence of harmful effects of redlining policies on NO$_2$ concentrations, with an estimated effect of 0.48 ppb (95\% CI: 0.35 to 0.62) for `B-A', 0.81 (0.65 to 0.96) for `C-A', and 0.87 (0.67 to 1.08) for `D-A', 
% Historically redlined regions are exposed to higher concentrations of NO$_2$ in 2010, suggesting that redlining policies contributed to these elevated air pollution concentrations 
even 36 years after the policy ended. 
We find evidence of weak harmful effects on PM$_{2.5}$ concentrations after adjusting for unmeasured confounding, with an estimated effect of 0.07 $\mu$g/m$^3$ (95\% CI: 0.03 to 0.11) for `B-A', 0.06 (0.01 to 0.10) for `C-A', and 0.04 (-0.02 to 0.10) for `D-A'. 
In addition, we can not dismiss the possibility that redlining once influenced PM$_{2.5}$ concentrations-an effect that may have diminished over time. These findings suggest that redlining has had a more pronounced impact on NO$_2$ concentrations. 

NO$_2$ and PM$_{2.5}$ pollutants originate from different sources (\cite{epano2}, \cite{epapm}). A potential explanation for the disparity in impacts between NO$_2$ and PM$_{2.5}$  may lie in highway vehicles, which is the primary contributor to NO$_2$. Highway vehicles could act as a mediating factor between redlining policies and NO$_2$ exposure. 

To explore the variance of causal effects, we revise the model to include random effects. For NO\(_2\), the population mean from the random effects model are 0.14 ppb (95\% CI: -0.04 to 0.31) for `B-A', 0.37 (0.16 to 0.57) for `C-A', and 0.50 (0.22 to 0.79) for `D-A'. Most cities present harmful effects, although only a few are statistically significant.
For PM$_{2.5}$, the population mean from the random effects model are 0.04 \(\mu g/m^3\) (95\% CI: -0.02 to 0.09) for `B-A', 0.03 (-0.03 to 0.09) for `C-A', and -0.01 (-0.08 to 0.07) for `D-A'. The harmful effects are predominantly concentrated along the East Coast and in certain Midwestern and Western cities. This pattern aligns well with the early urbanized areas of the 1930s and 1940s. Conversely, protective effects are more apparent in middle and northern cities, which largely urbanized during the Great Migration of the 1960s. 

To our knowledge, this is the first study to investigate the causal effect of redlining policies on air pollution concentrations and one of the earliest to explore the causal effect of redlining policies on environmental risk exposure. 
The key strengths of this study are the following: (1) we conduct exhaustive adjustment for potential unmeasured confounding. We adjust for city-level confounders, spatial confounders using spatial splines, and confounder of socio-economic status using proxy variables. (2) We prove the identification of causal effect given the data generating process modeled. (3) We quantify uncertainty using Bayesian MCMC. (4) We conduct intensive simulation study to demonstrate the performance on estimation and inference of our method over other currently available bias correction methods.

Our study has several limitations in data and modeling. First, the air pollution data are not fully observed; they are derived from empirical models \citep{kim2020concentrations}, and we do not account for this uncertainty. Second, our study covers only 69 cities. Historically, more cities were redlined \citep{nelson2023mapping}. %  provides maps of 202 cities
The covered 69 cities might be the most urbanized, considering that they are covered in the 1940 U.S. census while others are not. This suggests that our study may not be representative of the entire redlined population. 

%  The \backmatter command formats the subsequent headings so that they
%  are in the journal style.  Please keep this command in your document
%  in this position, right after the final section of the main part of 
%  the paper and right before the Acknowledgements, Supplementary Materials,
%  and References sections. 

% \backmatter

%  This section is optional.  Here is where you will want to cite
%  grants, people who helped with the paper, etc.  But keep it short!

\section*{Acknowledgements}
This research was supported by NIH-NIEHS grant 1R01ES031651. The authors thank Nate Wiecha for the help with data collection. \vspace*{-8pt}

%  If your paper refers to supplementary web material, then you MUST
%  include this section!!  See Instructions for Authors at the journal
%  website http://www.biometrics.tibs.org

\begin{singlespace}
\bibliographystyle{rss}
\bibliography{refs}
% \printbibliography  
\end{singlespace}

% \section*{Supplementary Materials}
% Web Appendices, Tables, Figures referenced are available with this paper at the Biometrics website on Oxford Academic.
% Please also describe the availability of data/code (see below) in this Supplementary Materials section, and/or if you refer to software found on a public repository such as Github. 

\label{lastpage}

\end{document}

\linenumbers

\begin{center}
{\Large Title of the paper}\\\vspace{6pt}
{\large Supplemental materials}\\
\today
\end{center}

\section{Proofs}\label{s:supp:proofs}

\section{Additional simulation results}\label{s:supp:sim}

\section{Additional data analysis}\label{s:supp:data}